\documentclass[12pt,preprint]{aastex}
\newcommand{\Msun}{\ifmmode\mbox{M}_{\odot}\else$\mbox{M}_{\odot}$\fi}
\newcommand{\Rsun}{\ifmmode\mbox{R}_{\odot}\else$\mbox{R}_{\odot}$\fi}
\newcommand{\Mearth}{\ifmmode\mbox{M}_{\oplus}\else$\mbox{M}_{\oplus}$\fi}
\newcommand{\Rearth}{\ifmmode\mbox{R}_{\oplus}\else$\mbox{R}_{\oplus}$\fi}

\shorttitle{Short Title Goes Here}
\shortauthors{Dib et al.}
\begin{document}
\title{{\emph{RXTE}} Observations of Anomalous X-ray Pulsar
1E~1547.0$-$5408 During and After its 2008 and 2009 Outbursts}
\author{Rim~Dib\altaffilmark{1},
        Victoria~M.~Kaspi\altaffilmark{1}, 
	Paul~Scholz\altaffilmark{1}, and
	Fotis~P.~Gavriil\altaffilmark{2,}\altaffilmark{3}}
\altaffiltext{1}{Department of Physics, McGill University,
                 Montreal, QC H3A~2T8}
\altaffiltext{2}{NASA Goddard Space Flight Center, 
Astrophysics Science Division, Code
662, Greenbelt, MD 20771}
\altaffiltext{3}{Center for Research and Exploration in Space Science
and Technology,
University of Maryland Baltimore
County, 1000 Hilltop Circle, Baltimore, MD 21250}
\begin{abstract}
We present the results of {\emph{Rossi X-ray Timing
Explorer}} ({\emph{RXTE}}) and {\emph{Swift}} monitoring observations of the magnetar
1E~1547.0$-$5408
following the pulsar's radiative outbursts in
2008~October and 2009~January. We report on a study of the evolution
of the timing properties and the pulsed flux from 2008 October~4 through 
2009 December~26. 
In our timing study, a phase-coherent
analysis shows that for the first 29 days following the 2008
outburst, there was a very fast increase in the magnitude of the
rotational frequency derivative $\dot{\nu}$, such that $\ddot{\nu}$
was a factor of $\sim$~60 larger than that reported in data
from 2007.  This $\dot{\nu}$ magnitude increase occurred in concert with the
decay of the pulsed flux following the start of the 2008 event.  Following the 2009
outburst, for the first 23 days, $\ddot{\nu}$ was consistent with
zero, and $\dot{\nu}$ had returned to close to its 2007 value. 
In contrast to the 2008 event, the 2009 outburst showed a major increase
in persistent flux, relatively little change in the pulsed flux, and sudden significant
spectral hardening $\sim$15~days after the outburst.
We show that, excluding the month following each of the outbursts, and
because of the noise and the sparsity in the data, multiple plausible
timing solutions fit the pulsar's frequency behavior.  We note
similarities in the behavior of 1E 1547.0$-$5408
following the 2008 outburst to that seen in the AXP 1E 1048.1$-$5937
following its 2001-2002 outburst and discuss this in terms of the
magnetar model.
\end{abstract}
\keywords{pulsars: individual(\objectname{1E~1547.0$-$5408}) ---
	  stars: neutron ---
 	  X-rays: stars}
\section{Introduction}

Anomalous X-ray Pulsars (AXPs) are young, isolated pulsars generally with a
large inferred magnetic fields ($\gtrsim$~10$^{14}$~G). They are detected
across the electromagnetic spectrum from the radio band (in 3 cases) to the
hard X-ray regime. Like a closely related class of pulsars, the
Soft Gamma Repeaters (SGRs), AXPs exhibit a wide range of variability,
including but not limited to flux changes, spectral variability, 
X-ray bursts, X-ray flares, pulse profile changes, timing glitches, 
and rapid changes in the rotational frequency derivative.
AXPs and SGRs are generally believed to be forms of magnetars. The magnetar
model \citep{td95,td96a,tlk02} identifies the power source of these
objects to be the decay of their strong magnetic fields. Most of the
known magnetars have been observed to have entered an active phase at least once in the past,
during which one or more of the listed types of
variability was observed. For recent reviews, see \cite{k07}, \cite{m08}, or
\cite{re11}.

The X-ray source 1E~1547.0$-$5408\footnote{Also known as SGR
1550$-$5418.} was discovered in 1980 with the {\emph{Einstein}} satellite during a
search for X-ray counterparts of unidentified X-ray sources
\citep{lm81}. It was first proposed to be a magnetar by \cite{gg07}
who showed that the spectrum was well described by a blackbody
plus power-law model, typical of AXPs, and that the source is located at the center of a
radio shell which is possibly a previously unidentified supernova
remnant.

Since its discovery, the source has exhibited many X-ray flux
variations \citep[see][for a review]{bis+11}.  First, a
comparison of the X-ray flux of 1E~1547.0$-$5408 between 1980 and 2006
observed by {\emph{Einstein}}, {\emph{ASCA}}, {\emph{XMM}}, and
{\emph{Chandra}} done by \cite{gg07} revealed that the source's
absorbed 0.5$-$10~keV X-ray flux decreased during this period by a
factor of $\sim$~7. Then, analysis of {\emph{Swift}} data taken in
2007 June and presented in \cite{crhr07} showed that the source's
X-ray flux was $\sim$3 times higher than the highest historic level
and $\sim$16 times higher than the lowest level.  

\cite{crhr07}
observed 1E~1547.0$-$5408 in the radio band and
detected pulsations at a period of $\sim$~2.069~s, the
smallest yet seen in a magnetar. The detected pulse profile was
much wider than those of most rotation-powered long-period pulsars,
and it was highly variable. Camilo et al. also estimated $\dot{P}$ to be
2$\times$10$^{-11}$, which, together with the value of $P$, imply a
magnetar strength field (assuming $B\equiv 3.2\times 10^{19}$~G$\sqrt{P
\dot{P}}$) of $2.2 \times 10^{14}$~G), 
and furthermore that this is the magnetar with the largest \footnote{See the online
magnetar catalog at www.physics.mcgill.ca/$\sim$pulsar/magnetar/main.html.}
$|\dot{\nu}|$ yet known.

Following the discovery of radio pulsations, \cite{hgr+08} studied the
X-ray emission of 1E~1547.0$-$5408 from 2007~June to 2007~October with
{\emph{Swift}} and {\emph{XMM}}. They showed that the X-ray flux
faded by about $\sim$~50\% from June to August. It subsequently
stabilized.  This suggests that an X-ray outburst occurred prior to
2007 June. Halpern et al. also detected faint X-ray pulsations for the first
time with {\emph{XMM}}. Moreover, during this period of time, a radio timing
analysis showed that the pulsar's
$\dot{\nu}$ had systematically decreased by $\sim$25\% \citep{crj+08}.

In 2008 October, 1E~1547.0$-$5408 entered a new active phase. The
{\emph{Swift}} BAT instrument detected several soft ($<$~100~keV)
bursts from the location of 1E~1547.0$-$5408 \citep{ier+10}.
An analysis of X-ray data from a variety of telescopes following the
burst showed that the source was $>$50 times brighter than the lowest
reported level and that the spectrum was significantly harder
\citep{ier+10,kgk+10,nkd+11}. The data further showed X-ray
pulsations and a frequency derivative more negative --and changing --
compared with the value reported in the radio data in 2007.


The source entered a second, even more active phase in 2009 January
\citep{mgw+09,kgk+10} during which a large number of soft gamma-ray bursts
was observed, as were well defined dust-scattering rings
\citep{tve+10}.  \citet{enm+10a} reported the discovery of a hard X-ray
tail just after the outburst using {\it Suzaku}, a result corroborated
by \citet{bis+11}.  A detailed analysis of the bursts detected with
{\emph{Swift}} after 2009 January outburst is presented by
\citet{sk11}, in addition to an analysis of the {\it Swift} persistent
and pulsed flux data taken immediately following the initial trigger,
showing that the source's flux rose for 6 hrs following the trigger. 

A comparison of the 2008 and 2009 events showed interesting
differences.  The former showed a large pulsed flux increase whereas
the latter, though having an enormous persistent flux increase, showed
only a modest pulsed flux increase.  \citet{nkd+11} pointed out, in
part using data reported on in this work, 
that the spectral variations in {\it Chandra}
observations of the two epochs were similar, in contrast to very
different $\dot{\nu}$ behaviors (see \S\ref{sec:timing}). They also noted a
significant pulsed fraction/flux anti-correlation. \citet{bis+11}
reported on data from {\it INTEGRAL, XMM-Newton}, and the {\it
Chandra X-ray Observatory}, finding similar $\dot{\nu}$, flux and spectral results.
They summarize the source's flux behavior since 1980 and argue it
shows three well-defined flux `states.'

Here we present a detailed analysis of {\emph{RXTE}} monitoring observations
of 1E~1547.0$-$5408 following its 2008 October and 2009 January
outbursts, supplemented by multiple {\emph{Swift}} observations obtained during the
same time period. We report the results of an in-depth analysis of the timing
behavior, pulsed flux changes, and pulse profile variations. Our
observations are described in Section~2. Our timing, pulse
morphology, and pulsed flux and hardness ratio analyses are presented,
respectively, in Sections~3, 4, and~5. 


\section{Observations and Pre-Analysis}
\label{sec:observations} 
\subsection{{\emph{RXTE}} Observations}

The results presented here were obtained using the proportional
counter array (PCA) on board {\em{RXTE}}. The PCA consists of an array
of five collimated xenon/methane multi-anode proportional counter
units (PCUs) operating in the 2$-$60 keV range, with a total effective
area of approximately 6500~cm$^2$ and a field of view of
$\sim$1$^{\circ}$ FWHM \citep{jsg+96}.

\begin{figure}
\centerline{\includegraphics[scale=.9]{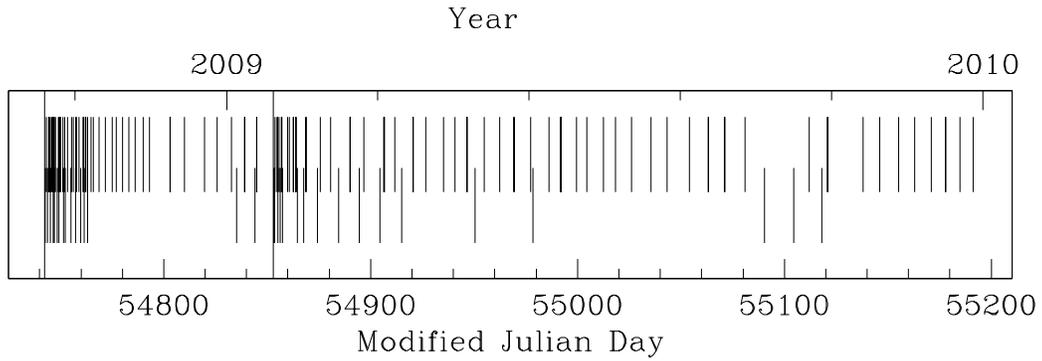}}
\caption{Epochs of the observations of 1E~1547.0$-$5408 used in this
paper. 
The first row of vertical bars marks the location of {\emph{RXTE}}
observations.
The second row of vertical bars marks the location of {\emph{Swift}}
observations.
The two solid vertical lines indicate the onset of the two outbursts.
\label{f1}}
\end{figure}

For this paper, we used 114 {\emph{RXTE}} observations. Of these, 49
observations were taken between 2008 October~4 and 2009 January~14,
following the 2008 October outburst, and 65 observations were taken
between 2009 January~22 and 2009 December~26, following the 2009
January outburst. The length of the observations varied between 0.6~ks
and 17~ks, but most were between 1~ks and 6~ks. The epochs of
the observations are shown in Figure~\ref{f1}.
Notice the high density of observations in the first few days
following each outburst.

Table~1 contains a list of the 49 observations obtained after the
2008~October outburst, showing the observation date, length, number of
PCUs on, the type of analysis that each observation was used for, and
the number of bursts found using the burst search algorithm introduced
in \citet{gkw02} and discussed further in \citet{gkw04}.  Table~2
contains a similar list of the 65 observations subsequent to the 2009~January
outburst.

For all observations, data were collected using the {\tt GoodXenon}
mode. This data mode records photon arrival times with 1-$\mu$s
resolution and bins photon energies into one of 256 channels. To
maximize the signal-to-noise ratio, we analysed only the events from
the top Xenon layer of each PCU. 

For each observation, we adjusted photon arrival times at each epoch
to the solar system barycenter, using the source position given in
\cite{gg07} ($\alpha$$_{J2000}$=15:50:54.11
$\delta$$_{J2000}$=$-$54:18:23.7). Then, for each observation, we
extracted a time series binned at a resolution of 1/32~s, in the
2$-$6.5~keV range (to maximize the signal to noise ratio), which
utilized all PCUs, to be used in the timing and pulse profile
analysis. Also for each observation, we extracted time series in
various energy ranges, binned at the same time resolution, but
excluding PCUs 0 and 1 because of the loss of their propane layers, to
be used for the pulsed flux, pulse profile, and hardness ratio
analyses. For each of the generated time series, 7-s long intervals of
data were removed around large bursts. We verified that 7-s was
sufficiently long. A large burst was defined to be any with count rate
greater than or equal to 50 counts~s$^{-1}$~PCU$^{-1}$ at the peak.

\subsection{{\emph{Swift}} Observations}

In this paper, {\emph{Swift}} observations were used uniquely for
timing purposes. We used all 16 observations taken between 2008
October~03 and 2009 January~12, 14 of which were taken in the month
following the onset of the 2008 outburst, and the remaining 2 taken in
2009 January, see Figure~\ref{f1}. We also used 8 {\emph{Swift}}
observations taken in the 23 days following the onset of the 2009
outburst, and all observations taken between 2009 February~13 and 2009
December~26 that were long enough to yield a clear pulse profile.
Table~3 contains a list of the {\emph{Swift}}
observations used in this work, all of which were in Windowed Timing (WT) mode.

For each observation, we extracted photons from a 40-pixel-long strip
centered on the source, and we adjusted the photon arrival times of
each epoch to the solar system barycenter. Then, for each observation,
we extracted a time series binned at a resolution of 1/32 s in the
2$-$7~keV range, to match the {\emph{RXTE}} band.

\section{Phase-Coherent Timing}
\label{sec:timing}

Time series for all 95 {\emph{RXTE}} observations having a sufficiently high signal-to-noise ratio
were divided in half and treated as two separate time series. Time
series for all 32 {\emph{Swift}} observations were not divided.
Each time series was epoch-folded using an ephemeris determined
iteratively by maintaining phase coherence as we describe below. When
an ephemeris was not available, we folded the time series using a
frequency obtained from a periodogram. Resulting pulse profiles, with
64 phase bins, were cross-correlated in the Fourier domain with a high
signal-to-noise-ratio template created by adding phase-aligned profiles. The
cross-correlation returned an average pulse time-of-arrival (TOA) for
each observation, corresponding to a fixed pulse phase.

The pulse
phase $\phi$ at any time $t$ can usually be expressed as a Taylor
expansion,

\begin{equation}
\label{eq:polynomials}
\phi(t) = \phi_{0}(t_{0})+\nu_{0}(t-t_{0})+
\frac{1}{2}\dot{\nu_{0}}(t-t_{0})^{2}
+\frac{1}{6}\ddot{\nu_{0}}(t-t_{0})^{3}+{\ldots},
\end{equation}

\noindent where $\nu$~$\equiv$~1/$P$ is the pulse frequency,
$\dot{\nu}$~$\equiv$~$d\nu$/$dt$, etc$.$, and subscript ``0'' denotes a
parameter evaluated at the reference epoch $t=t_0$. 

To obtain ephemerides for the time periods following each outburst, we
fitted the TOAs to the above polynomial using the pulsar timing
software package TEMPO\footnote{See
http://www.atnf.csiro.au/research/pulsar/tempo.}.  Since the spin-down
of this source was unstable, and since consecutive observations were
not always sufficiently close to each other, phase coherence could
only be unambiguously maintained for a period of 29 days following the first
outburst, and 23 days following the second outburst. The two timing
solutions found are represented by solid red lines in the first two
panels of Figure~\ref{f2}, along with the radio timing parameters of
the same source from 2007 \citep{crhr07,crj+08}. The spin parameters
for these two timing solutions are shown in Table~4. 
The 2008 and 2009 ephemerides are consistent
within 2$\sigma$ with those reported in
\cite{ier+10} and in \cite{bis+11}, although they extend for a few days
longer than in the previous analyses.

We made various attempts at finding the pulsar frequency outside of
these time periods. First, for each instance where closely
spaced {\emph{RXTE}} observations were available, a frequency measurement was found
by doing TEMPO fits through the closely spaced TOAs.  
These frequency measurements are shown as dark blue points in
the first panel of Figure~\ref{f2}.

Then, all 29 observations longer than 5.5~ks were broken into 2 or 3
fragments, TOAs were extracted for each fragment, and a frequency
measurement was found for each long observation by fitting a linear
trend to the phases with TEMPO. These frequency measurements are shown
as light blue points in the first panel of Figure~\ref{f2}.

We also made several attempts at finding phase-coherent timing
solutions outside the one month following each outburst. We broke the
list of all TOAs (one or two TOAs per observation) into 69 overlapping
sets of TOAs. For each set of TOAs, we searched for a phase-coherent
timing solution by running a TEMPO fit for every plausible combination
of $\nu$ and $\dot{\nu}$. The results of the search of one such set
of TOAs are presented in Figure~\ref{f3}.  For each guess combination
of $\nu$ and $\dot{\nu}$, TEMPO returned a best-fit $\nu$ and
$\dot{\nu}$, and timing residuals from which a reduced ${\chi}^2$ was
extracted and represented as a shade in Figure~\ref{f3}, with the
darker shade corresponding to a smaller reduced $\chi^2$. Note the
appearance of several ``islands'' on the Figure. The center of each
island corresponds to the best-fit $\nu$ and $\dot{\nu}$ returned by
TEMPO.


The fact that many islands appear in the Figure indicates that any
timing solution found to cover the time period spanned by the set of
TOAs is not unique. Instead, the various islands correspond to several
plausible timing solutions, differing from each other by a small
number of pulse phases. 

Once a shaded $\nu$ and $\dot{\nu}$ map, like that shown in
Figure~\ref{f3}, was obtained for each overlapping set of TOAs, we
collapsed each map, in the horizontal direction, into a single shaded
column. The shade at each height of the column corresponds to the
darkest shade in the corresponding row of the original $\nu$ and
$\dot{\nu}$ map before the collapse. We then plotted all the shaded
columns in Figure~\ref{f4}, along with the various phase-connected
timing solutions and along with the individual frequency measurements.

For the columns near the outbursts, the number of dark spots in each
shaded column was small, indicating a small number of possible
phase-connected timing solutions. However, the fact that the number of
dark regions in each shaded column increased as we got further from
the outbursts clearly indicates that timing solutions found outside of
the time periods covering the month following each outburst are not
unique. Some of the longest-running non-unique solutions are shown as
dotted red lines in Figures~\ref{f2} and ~\ref{f4}. In the 2008 event,
the loss of phase-coherence is almost certainly due to enhanced timing
noise in the pulsar as the frequency of observations had not
significantly dropped when phase coherence was lost; in the 2009
event, the time interval between observations was somewhat larger when
phase coherence was lost, although the coincidence of the loss and the
decay of the persistent flux suggests the source behavior plays an
important role.


\section{Pulse Profile Analysis}
\label{sec:profiles}

First, for each {\emph{RXTE}} observation, we folded the data in the 2--6.5~keV band
using the best-fit frequency found in the timing analysis (Section~3).  
We then cross-correlated the resulting profile with
a standard template in order to obtain phase-aligned profiles.

We then divided the period following the first outburst (3~October to
22~January) into intervals with similar net exposures, shown with the
letters A through K at the bottom of Figure~\ref{f5}. We did the same for the
period following the second outburst (22~January to 26~December~26); the
intervals are shown with the letters L through P at the bottom of Figure~\ref{f5}.

For each time interval, we summed the aligned profiles, subtracted the
DC component, and scaled the resulting profile so that the value of
the highest bin is unity and the lowest point is zero. The results are
presented in Figure~\ref{f5} with the time intervals marked in the top
left corner of each profile. The different profile qualities are due
to the changes in the pulsed flux of the pulsar. Nevertheless, clear
changes in pulse morphology are apparent.  For example, the profile
peak is clearly narrower immediately following the second 2008 pulsed
flux enhancement, and shows the full width, but is more flat-topped
following the 2009 outburst. The results shown here are generally
consistent with those of \citet{nkd+11} and \citet{bis+11}.

Note however that in no case are any pulse profile changes so large
that the profile peak, the assumed fiducial point, cannot be
unambiguously identified.  This is in contrast to other examples of
AXP outbursts in which multiple peaks appeared, rendering unambiguous
pulse numbering problematic \cite[e.g.][]{wkga11}. This gives us
confidence that the timing analysis reported in the previous Section,
in particular the difficulties phase-connecting the timing data, are
not significantly affected by pulse profile changes.
Indeed the epochs in which the profile changes are largest (namely
immediately following the outbursts) are those which we have
successfully phase-connected.

To study the evolution of the pulse profile with energy, we repeated
the above analysis for energy bands 2$-$4~keV, 4$-$10~keV, and
2$-$10~keV. Figure~\ref{f6} shows the evolution of the pulsed flux
with energy for time intervals M and~N indicated at the bottom of
Figure~\ref{f5}. Note that the location of the peak of the profile
seems to be energy-dependent, with the peak more to the left at higher
energies, and more to the right at lower energies.
This implies that spectral changes might cause slight shifts in the
location of the pulse peak in the energy band used for the timing
analysis (2$-$6.5~keV), but again, not large enough to cause a loss in
pulse counting.
Also note that
pulse profiles at higher energies are not shown because of the very
poor signal-to-noise ratio at these energies.

\section{Pulsed Flux Time Series and Hardness Ratios}
\label{sec:flux}

To obtain the pulsed flux for 1E~1547.0$-$5408, for each observation,
we folded the data and extracted aligned pulse profiles in several
energy bands.  For each folded profile, we calculated the RMS pulsed
flux,

\begin{equation}
F_{RMS} =
{\sqrt{2 {\sum_{k=1}^{n}}
(({a_k}^2+{b_k}^2)-({\sigma_{a_k}}^2+{\sigma_{b_k}}^2))}},
\label{eq:f1}
\end{equation}
\noindent where $a_k$ is the $k^{\textrm{\small{th}}}$ even Fourier
component defined as $a_k$ = $\frac{1}{N} {\sum_{i=1}^{N}} {p_i} \cos {(2\pi
k i/N})$, ${\sigma_{a_k}}^2$ is the variance of $a_k$, $b_k$ is the odd
$k^{\textrm{\small{th}}}$ Fourier component defined as $b_k$ = $\frac{1}{N}
{\sum_{i=1}^{N}} {p_i} \sin {(2\pi k i/N})$, ${\sigma_{b_k}}^2$ is the
variance of $b_k$, $i$ refers to the phase bin, $N$ is the total number of
phase bins, $p_i$ is the count rate in the $i^{\textrm{\small{th}}}$ phase
bin of the pulse profile, and $n$ is the maximum number of Fourier harmonics
used; here $n$=4. 

In the top panel of Figure~\ref{f7}, we show the RMS pulsed flux in
2$-$10~keV, 2$-$4~keV, and in 4$-$10~keV. Note the 2$-$10~keV time series was
previously shown by Ng et al. (2011) and Scholz \& Kaspi (2011).  The panel shows a slow decrease in
the pulsed flux after the first outburst interrupted by a spike 9 days after
the outburst. The spike was followed by a slow decrease.  The source behaves
differently after the second outburst, when the persistent flux increases
dramatically (Figure~\ref{f2}d) while the pulsed flux does not vary
significantly, implying a large decrease in the pulsed fraction following the
second outburst.  This is consistent with results of \citet{nkd+11},
\cite{bis+11}, and \citet{sk11}.

In the bottom panel of Figure~\ref{f7}, we show the hardness ratio as a
function of time, computed from ratio of the pulsed fluxes in the energy ranges
4$-$20~keV to 2$-$4~keV. The Figure shows a clear discontinuity in the pulsed
hardness ratio $\sim$~15 days after the outburst, in the time period
bracketed by the last two dotted lines, which, notably, is subsequent to the bulk
of the pulsed flux decay, but while the persistent flux was still relatively high. 
We also computed the hardness ratio
in various other energy bands, but the discontinuity was most significant
for the energy range shown.  Unfortunately {\it Suzaku} \citep{enm+10a} and {\it INTEGRAL} \citep{bis+11}
hard X-ray observations of the source both occurred prior to the hardness change seen
with {\it RXTE}.

\section{Discussion}
\label{sec:discussion}





The behavior of 1E~1547.0$-$5408 over the past several years has been
extremely complex, indeed mirroring the complexity of the behaviors of
AXPs in outburst in general.  In a collection of detailed analyses of
various X-ray and gamma-ray data sets
\citep{ier+10,kgk+10,nkd+11,sk11,bis+11}, including the present {\it RXTE}
study reported on herein, a picture emerges of dramatic and varied
multiwavelength behavior, including at least two very different
outbursts from this intriguing source, albeit bearing some common
properties.  In the latter category is the generic X-ray hardness/flux
correlation seen now commonly in magnetar outbursts
\citep[see][for reviews]{re11,sk11}, X-ray pulse profile changes, and
ubiquitous X-ray bursting near outburst epochs, and significant
changes in spin-down rate at and surrounding outburst epochs.
We focus on this latter point here.

Magnetar rotational evolution during and near radiative outbursts has shown
a wide range of behaviours, including
large spin-up glitches with unusual recoveries \citep[e.g.][]{kgw+03},
likely spin-down events \citep[e.g.][]{pal02,wkg+02}, initial spin-up events
paired with large over-recovery, yielding net spin-down \citep{lkg10},
and large $\dot{\nu}$ variations on week-to-month time scales
\citep{gk04,dkg09}. We note that throughout, no AXP has ever been
observed to spin {\it up} for an extended period of any duration; only
changes in magnitude in the spin-{\emph{down}} rate punctuated by
occasional, sudden, unresolved spin-up events have yet been observed. This
is in stark contrast to behavior generally seen in accreting systems
\citep[e.g.][]{cbf+97}.  Nevertheless, the diversity in magnetar
rotational evolutions near the epochs of radiative outbursts has been
perplexing.

However, we note that the behavior of 1E 1547.0$-$5408
following its 2008 outburst is reminiscent of unusual
timing evolution seen thus far uniquely in AXP 1E~1048.1$-$5937 following 
its 2002 flux `flare' \citep{gk04,dkg09}.  We compare the two next.

In the case of 1E~1048.1$-$5937, the pulsar's pulsed flux had been
relatively stable in several years of {\it RXTE} monitoring, when a
small (factor of $\sim$2) pulsed flux increase was observed near the
end of 2001 (see Fig. 8e), accompanied by short SGR-like bursts \citep{gkw02} and
pulse profile changes \citep{dkg09}. The pulsed flux relaxed back to
near quiescence in $\sim$3 weeks, but then there was a second, larger
pulsed flux increase, this time by a factor of $\sim$3 over the span
of a few weeks.  This larger pulsed flux `flare' subsequently decayed
on a time scale of several months.  This latter decay occurred
just prior to an increase in the magnitude
of $\dot{\nu}$, by over an order of magnitude (Fig. 8c). During both flares, the
pulsar suffered significant pulse profile changes, the largest
of which were at the flare peak, with the pulse relaxing back on
approximately the same time scale as the pulsed flux dropped.
Subsequently, there was a $\sim$2-yr period of large
variations in $\dot{\nu}$, which while always remaining negative,
varied by as much as an order of magnitude on time scales
of a few weeks (see Fig.~15 in \citealt{dkg09}; Fig. 8c).  Phase connection in
this extremely noisy interval was difficult, but enabled by extremely
frequent {\it RXTE} observations -- approximately three per 7--10-day
interval. 

The 2008 outburst of 1E 1547.0$-$5408 showed similar
behavior (Fig. 8). The pulsed flux was clearly enhanced in our {\it RXTE} observations,
which were triggered by SGR-like bursts. 
This enhancement then started to relax, but the decay was interrupted
by a second pulsed flux increase $\sim$2 weeks later, this time by a
factor of $\sim$4, similar to the increase in the pulsed flux in the
second 2002 flare of 1E 1048.1$-$5937 (compare Fig. 8e and f). 
The pulse profile changed,
particularly near the peak of the second `flare.'
Moreover, $\dot{\nu}$, as for
1E~1048.1$-$5937, dropped by a factor of $\sim$4,
but roughly in concert with the pulsed flux decrease,
before phase-coherence was lost (Fig. 8d).
It is plausible that large $\dot{\nu}$ variations ensued, but we cannot
unambiguously confirm this as the spacing of subsequent {\it RXTE}
and {\emph{Swift}} observations was not as dense as for 1E 1048.1$-$5937.
Clearly though, $\dot{\nu}$ recovered to its 2007 value by the time of
the 2009 outburst, so another dramatic change must have occurred.  

Thus, the behavior of 1E 1547.0$-$5408 during its 2008 outburst is
similar to that of 1E 1048.1$-$5937 in 2001-2002, as suggested in
Figure 8: in both cases
there was an initial, smaller flare accompanied by bursting, and a
much larger flare a few weeks later. In both cases, the flares were
accompanied by pulse profile changes in which the largest changes occurred
near the peaks of the flares\footnote{We note that the 2007 flare of 1E 1048.1$-$5937
also showed a double-peaked pulsed flux variation accompanied by significant pulse
profile changes; see Fig. 15 in \citet{dkg09}.}. Then, importantly, a pulsed flux decay
occurred near or simultaneous with an increase in the magnitude of
$\dot{\nu}$ by a factor of $\sim$4, followed by a very noisy period, in which
regular phase-coherent timing was impossible.  We note these similarities
exist in spite of the fact that 1E 1048.1$-$5937 is a persistent, bright AXP,
while 1E~1547.0$-$5408 has the properties of a transient AXP.

Note however that this behavior was not replicated in the 2009 1E 1547.0$-$5408
outburst: in that case $\dot{\nu}$ remained close to that reported in
the radio data by \citet{crhr07}.  Also, in the 2009 outburst, the
pulsed flux was barely affected, while the persistent flux increased
by an order of magnitude (Fig. 2).  Still, phase-coherence
became again impossible some 23 days after the 2009 outburst. We note
low-level but significant pulsed flux variations subsequently, which
could be relevant.

The similarity between the 2008 outburst of 1E~1547.0$-$5408 and the
2001-2002 outburst of 1E~1048.1$-$5937 is potentially interesting.  In the magnetar
model, large flux enhancements are easy to envision given the great
stresses placed on the crust due to internal magnetic field decay.
Moreover, field-line distortions in the magnetosphere are expected as
the crust shifts and the magnetosphere twists, hence pulse profile
changes are also generically expected.  \citet{bel09} argues that
magnetospheric twists will impact the pulsar spin-down in two possible
ways. One is that a strong twist results in an enhancement of the
poloidal field lines, which leads to a stronger spin-down torque on
the star \citep{tlk02}.  This enhanced $\dot{\nu}$ is then expected to
decay back to the standard value.  The second way is that once the
twist angle $\psi$ reaches a maximum value, it ``boils over,'' with
energy that would have been stored in the magnetosphere had $\psi$
continued to increase being carried away by an intermittent magnetic
outflow, which can carry away significant angular momentum. In this
picture, any delay between such intermittent magnetic outflows and the
outburst onset is due to the time it takes $\psi$ to reach its maximal
value, though the increase in $\dot{\nu}$ is expected to commence much
earlier \citep{bel09}.

While the torque variations picture put forth by \citet{bel09} may
provide a qualitative explanation of the behavior we have seen in the
2001-2002 outburst of 1E 1048.1$-$5937 and the 2008 outburst of 1E
1547.0$-$5408, it remains to be seen if the model can explain
quantitatively the large $\dot{\nu}$ variations observed unambiguously
for 1E 1048.1$-$5937, namely order-of-magnitude magnitude changes.
Indeed the magnitude of the initial torque increase for strong ($\psi
> 1$) twists was not estimated by Beloborodov (2009), as it requires a full nonlinear
calculation. An initial estimate for moderate twists suggests the
fractional change in effective dipole field due to poloidal inflation
varies as $\psi^2$, which could be checked if many similar outbursts
are observed.

However, in the case of 1E~1547.0$-$5408, it is difficult to
understand why the magnitude of $\dot{\nu}$ remained small and
unchanged in the 23 days following the 2009 event trigger, while the
persistent X-ray luminosity decayed from an order-of-magnitude
increase.  That there was very little change in the pulsed flux seems
relevant, suggesting any poloidal field enhancement has to be
accompanied by a pulsed flux increase.  This seems reasonable if the
pulsed emission emerges from the polar region, a hypothesis that may be testable
via X-ray polarimetry. Perhaps then the 2009
event involved a much greater fraction of the neutron star, and, at
least initially, impacted the poloidal region minimally.

As discussed by \citet{gk04}, disk models appear to have difficulty
explaining the changes in pulsed flux and torques we have observed.  
This is because, at least for the propeller
torque prescription of \citet{chn00}, the X-ray luminosity is expected
to vary as $\dot{\nu}^{7/3}$, which is not consistent with the data.
Moreover, the disparate behaviors in 2008 and 2009 also pose a major
challenge to any disk model, though are also not obviously explained
in the \citet{bel09} picture.


The long-term timing behavior of the source is summarized in Figures
2(a) and (b).  Clearly the enhancement in the magnitude of $\dot{\nu}$
is an interesting event in the pulsar's long-term spin-down history.
The early 2007-2008 radio data (shown in green in Fig. 2a) show that a
quadratic polynomial was required to describe the spin-down
\citep{crhr07,crj+08}, with $\ddot{\nu}$ $\sim$60 times smaller than
that seen in 2008.  If the radio ephemeris is extrapolated to the
start of the 2008 outburst, it over-predicts the frequency by
approximately $\sim$30~$\mu$s, consistent with there having been other
episodes of enhanced spin-down. If these were small in number, we
might have expected accompanying radiative events as in 2008, but none
are reported. This supports the 2008 enhanced spin-down being
particularly notable.

\begin{figure}
\centerline{\includegraphics[scale=.67]{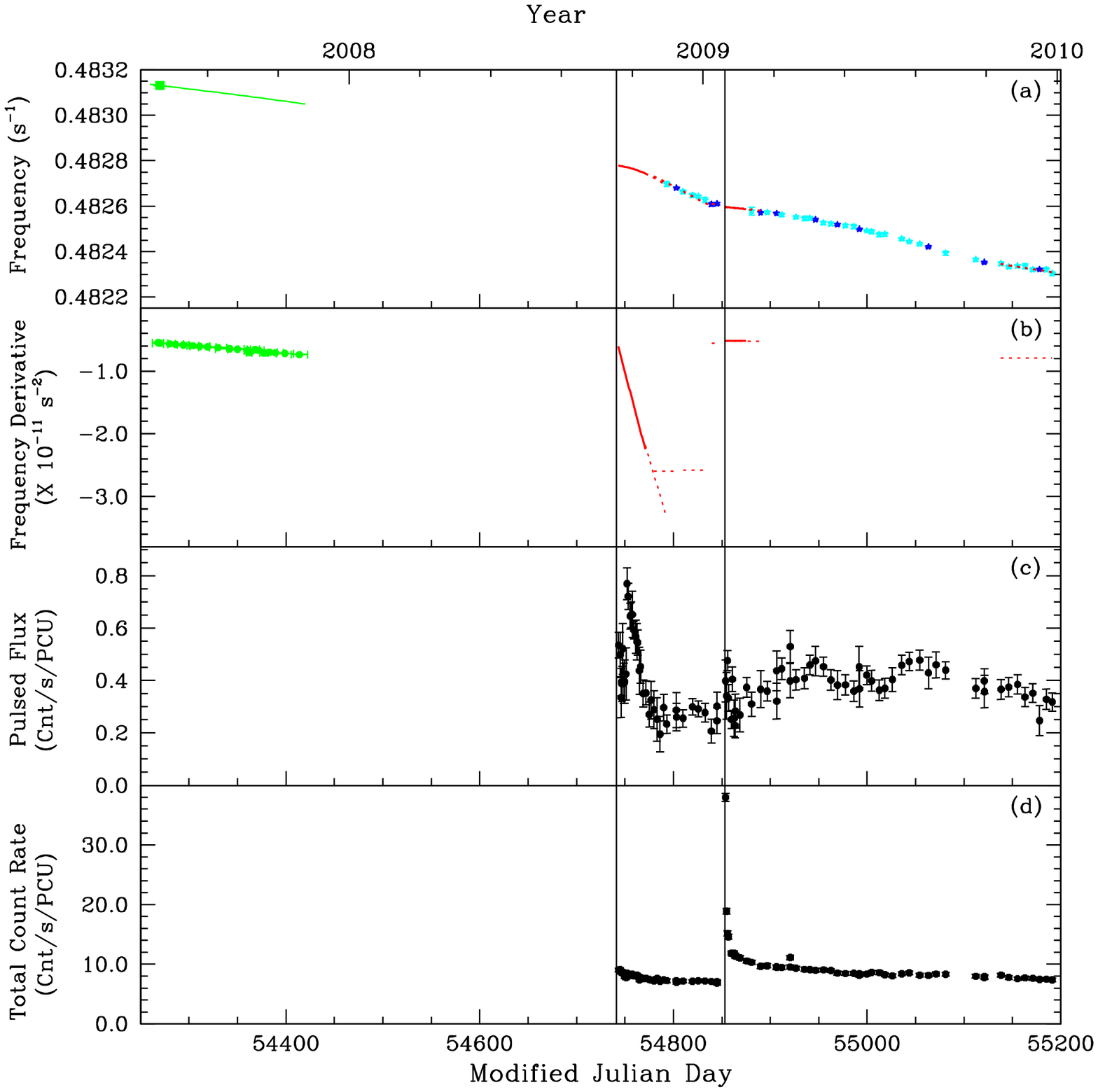}}
\caption{Long-term evolution of 1E~1547.0$-$5408 properties.
(a) Frequency as a function of time. Green square: frequency from
radio data \citep{crhr07}. 
Green line: evolution of the frequency
$\nu$ as a quadratic polynomial where ${\nu}_o$ and ${\dot{\nu}}_o$
are obtained from \cite{crhr07}, and ${\ddot{\nu}}$ is from the
best-fit slope to Panel~2 of Figure~6 in \cite{crj+08}. 
Red solid lines: frequency as a function of time from phase-coherent timing. 
Red dotted lines: possible but not unique phase-coherent solutions. 
Dark blue points: individual frequency measurements, each obtained from a
combination of closely spaced observations. 
Light blue points: individual frequency measurements, each obtained from
a long observation.
(b) Frequency derivative as a function of time.
Green points: $\dot{\nu}$ measurements taken from \cite{crj+08}.
Red solid lines: $\dot{\nu}$ as a function of time from phase-coherent
timing.
Red dotted lines: possible but not unique phase-coherent solutions.
(c) Pulsed flux in the 2$-$10~keV band.
(d) Count rate in the 2$-$10~keV band from the full {\emph{RXTE}}
field of view.
All panels: the two solid vertical lines mark the onset of the two
outbursts.
\label{f2}}
\end{figure}
\begin{figure}
\includegraphics[scale=.8]{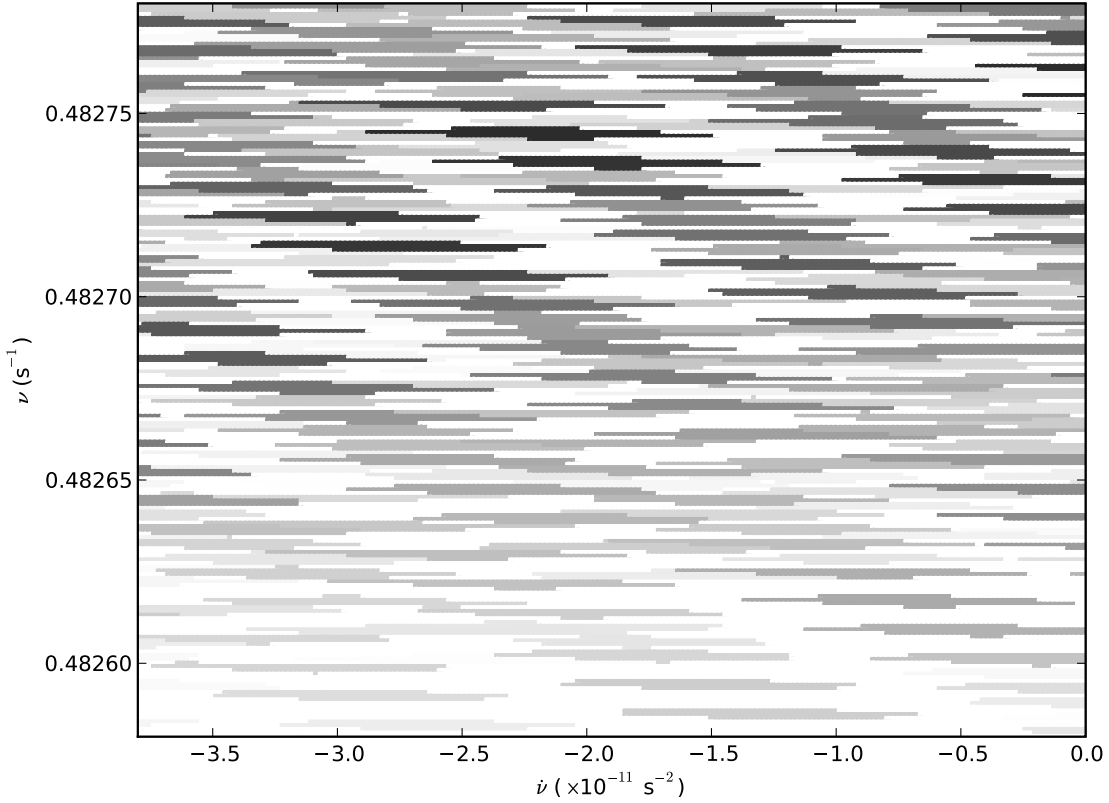}
\caption{Results of the
search for a phase-coherent timing solution consisting of a frequency
$\nu$ and of a single frequency derivative $\dot{\nu}$ for a set of
TOAs covering a time period of 11 days, and starting 21 days after the
onset of the first outburst (MJD 54764.846 to 54775.050).  The x-axis
shows trial $\dot{\nu}$ values.  The y-axis shows trial $\nu$ values.
For every pair of $\nu$ and $\dot{\nu}$, the shading corresponds to
the reduced ${\chi}^2$ obtained when the pair was used as set of trial
phase-coherent timing parameters by TEMPO. White squares correspond to
a reduced ${\chi}^2$ of 6 and above. Black squares correspond to a
reduced ${\chi}^2$ of 0. The center of each of the uniformly-shaded
islands corresponds to the best-fit $\nu$ and $\dot{\nu}$ found by
TEMPO for each point on that island, and the size of each island
corresponds to the uncertainty on the obtained $\nu$ and $\dot{\nu}$.
The presence of several black islands clearly indicates the presence
of several plausible timing solutions differing from each other by a
small number of phase counts.
\label{f3}}
\end{figure}
\begin{figure}
\includegraphics[scale=.72]{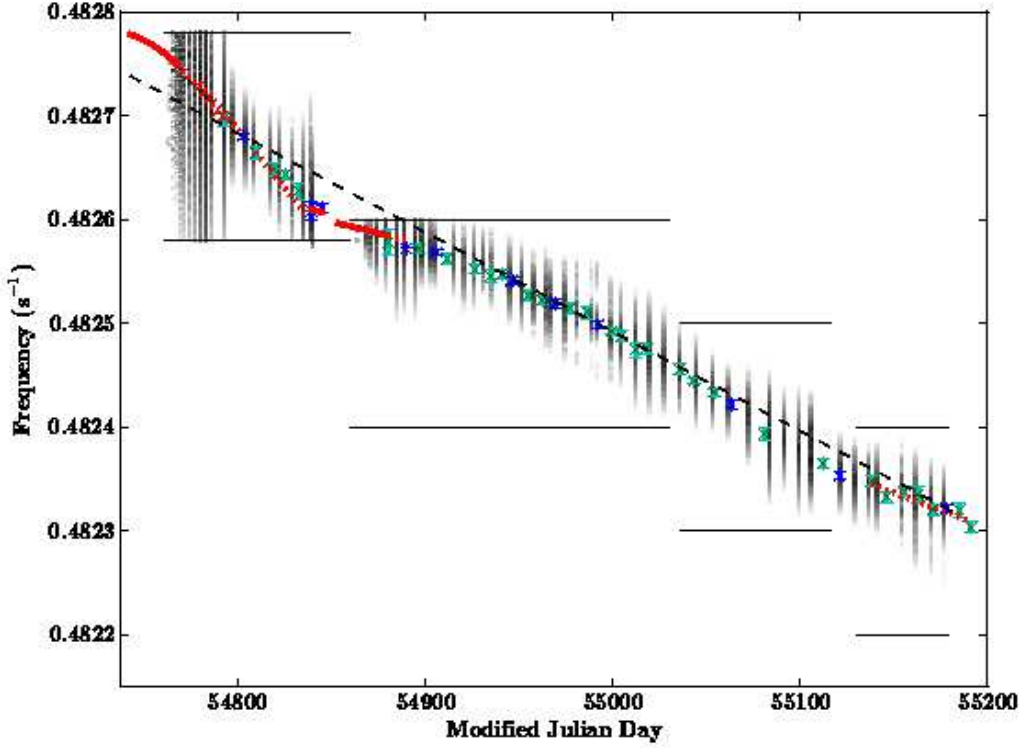}
\caption{
Frequency as a function of time for 1E~1547.0$-$5408.
Shaded black columns: each of the black columns corresponds to one
of the overlapping sets of TOAs. The x-coordinate of the column
corresponds to the time in the middle of that set of TOAs.
Every column was obtained by collapsing vertically a rectangle
similar to that shown in Figure~\ref{f3}.
The shading of each square in every shaded column appearing in
this Figure corresponds to the darkest shading in the corresponding
row in a $\nu$ and $\dot{\nu}$ map similar to that shown in
Figure~\ref{f3},
and represents the best-fit reduced ${\chi}^2$.
The presence of several black squares in every column
indicates the presence of several plausible phase-coherent timing
solutions at each epoch.
The black horizontal lines represent the limits of the searched
frequency values.
The red solid lines represent frequency as a function of time 
from phase-coherent timing.
The red dotted lines represent possible but not unique 
phase-coherent solutions.
The dark blue points represent individual frequency measurements, 
each obtained from a set of closely spaced observations.
The light blue points represent individual frequency measurements, 
each obtained from a long observation.
The dashed line is an eyeball fit to all frequency measurements
from 2008 and 2009.
The two vertical black lines denote the onset of the two outbursts.
\label{f4}}
\end{figure}
\begin{figure}
\includegraphics[scale=.8]{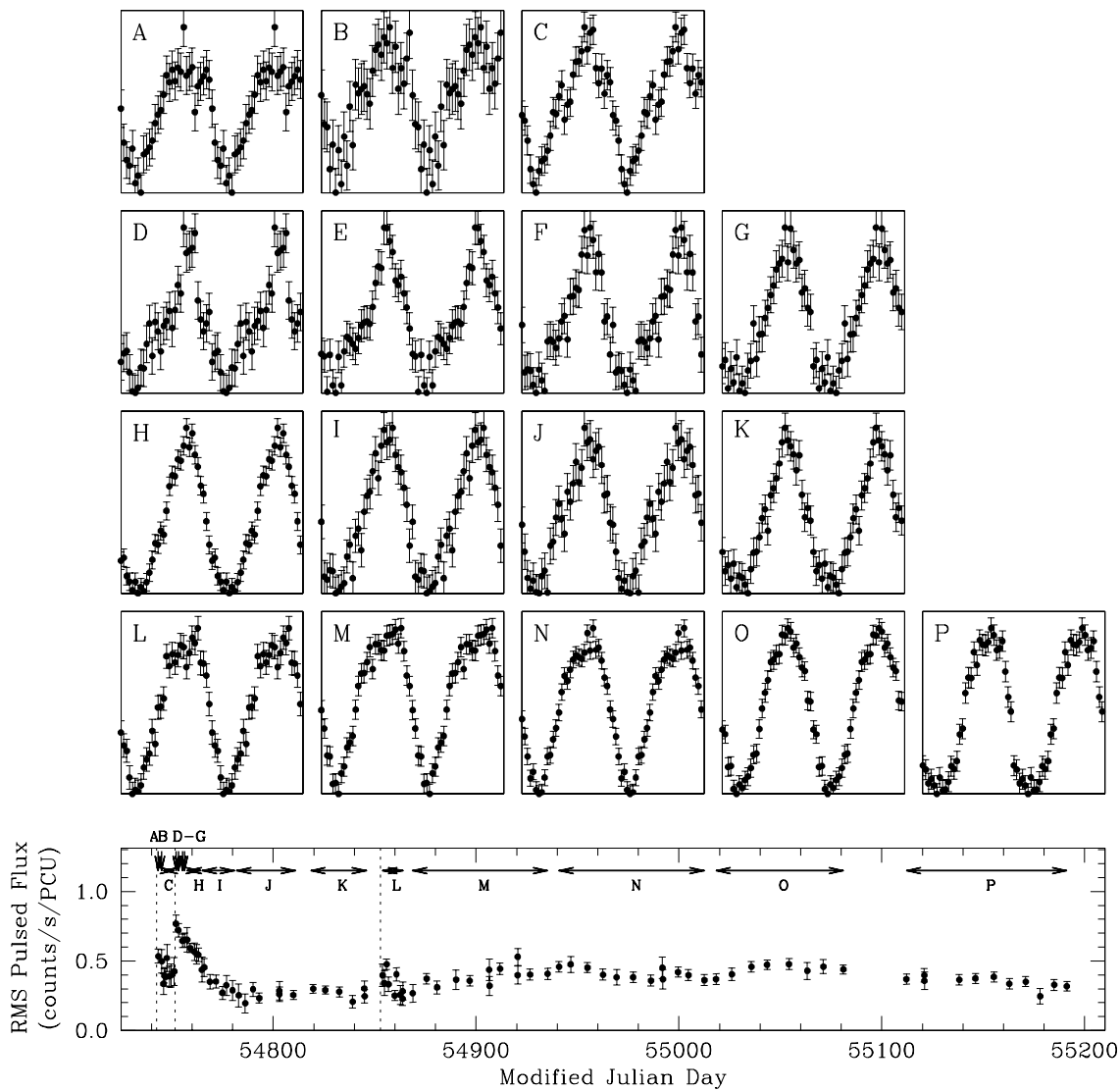}
\caption{Normalized 2$-$6.5 keV pulse profiles of 1E~1547.0$-$5408 from
2008 October to 2010 December. Two cycles per plot are shown for clarity.
The letter shown in the top-left corner
of each plot refers to the time segments marked by arrows in the
bottom plot, where the 2$-$10 keV RMS pulsed flux is shown for
reference. The profiles marked with the letters A, B, and D to G
are obtained from individual observations. 
The first and third dotted
lines in the bottom plot correspond to the onsets of the two outbursts.
\label{f5}}
\end{figure}
\begin{figure}
\centerline{\includegraphics[scale=.8]{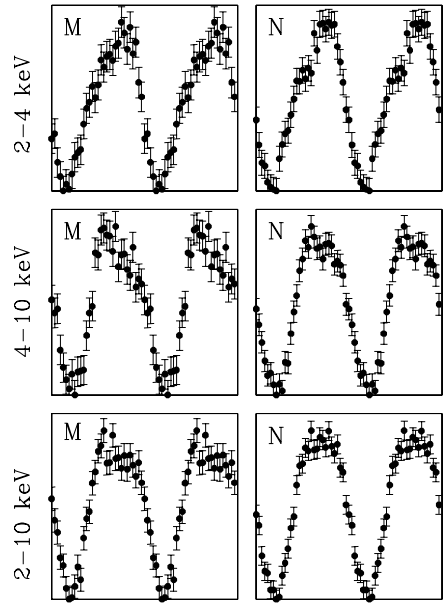}}
\caption{Normalized 2$-$4 keV, 4$-$10~keV, and 2$-$10~keV pulse
profiles of 1E~1547.0$-$5408 for two selected time segments
showing the dependence of the location of the peak on energy.
Two cycles per plot are shown for clarity.
\label{f6}}
\end{figure}
\begin{figure}
\includegraphics[scale=.8]{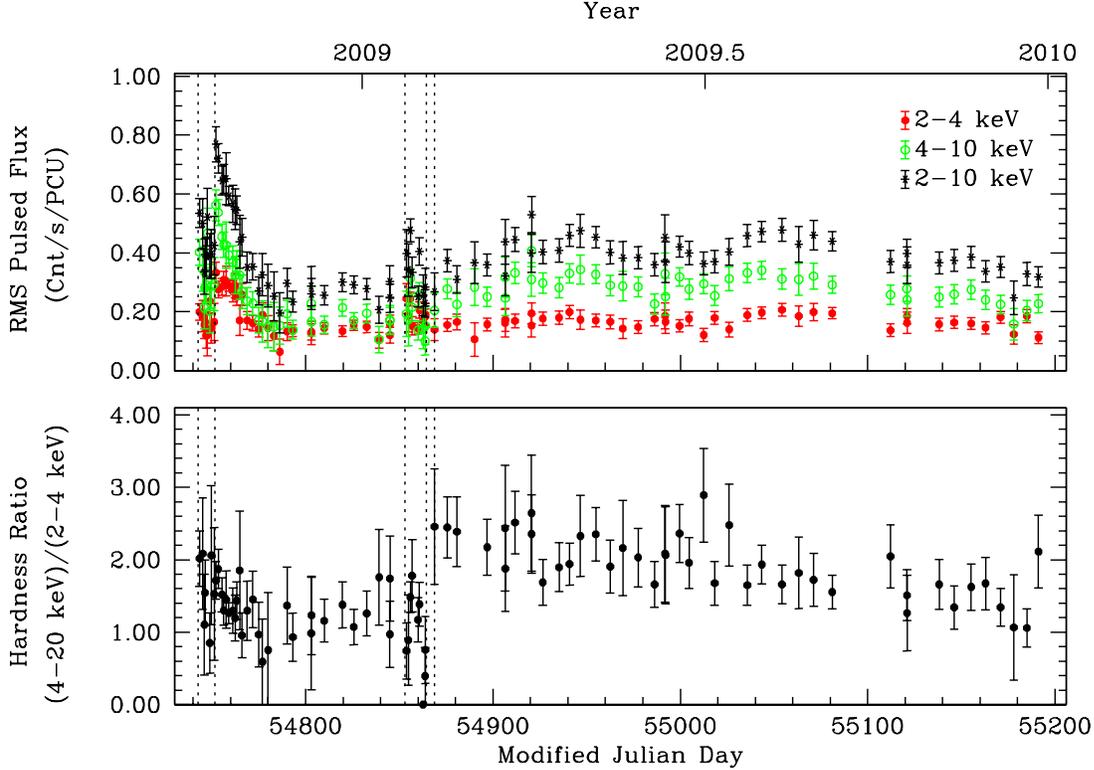}
\caption{Pulsed flux time series and pulsed hardness ratio of
1E~1547.0$-$5408. 
Top: evolution of the RMS pulsed flux of 1E~1547.0$-$5408
in 2$-$4~keV, 4$-$10~keV, and 2$-$10~keV.  Note the latter time
series was shown by Ng et al. (2011) and Scholz \& Kaspi (2011).
Bottom: The hardness ratio as a function of time, computed 
from the ratio of pulsed fluxes in the energy range
4$-$20~keV to 2$-$4~keV. All panels:
The first dotted line marks the onset of the first outburst.
The second dotted line marks the location of an enhancement in the
pulsed flux. The third line marks the onset of the second outburst. 
The last two dotted lines bracket the time interval inside which 
the hardness ratio rose.
\label{f7}}
\end{figure}

It was noted in \cite{dkg09} that every observed AXP flare or outburst
thus far were accompanied by a timing event, possibly caused by some
unpinning of fluid vortices, which in turn was caused by crustal
movement due to a twist propagating outward. \cite{dkg09} also noted
that the converse is not true: many AXP glitches appear to be
relatively silent. This leads to the question of whether the onsets of
the 2008 and 2009 outbursts of 1E~1547.0$-$5408 were accompanied by
glitches. For the 2008 outburst, this question is impossible to answer
given the lack of observations prior to the outburst. The question is
also difficult to answer for the 2009 outburst given the sparsity of
data and loss of phase coherence prior to the outburst. To provide
an answer, we found a phase-coherent solution covering
only the three observations just prior to the onset of the 2009 outburst (2
{\emph{RXTE}} observations and one {\emph{Swift}} observation,
ephemeris covering MJD 54839$-$54845, see the small red lines just
prior to the 2009 oubutst in Panels~1 and~2 of Figure~\ref{f2}). This
returned the following parameters for reference epoch 54845.0:
${\nu}$=0.4826070(3)~s$^{-1}$,
$\dot{\nu}$=$-$5.6(1.3)$\times$10$^{-12}$~s$^{-2}$. When this solution
is extrapolated forward in time to the date of the first post-onset
observation, the frequency obtained is
9.6(5)$\times$10$^{-6}$~s$^{-1}$ {\emph{larger}} than the known
frequency at that epoch. This change in frequency is in a direction
opposite to that which a spin-up glitch would cause.
This does not rule out the possibility of a glitch, because a glitch
followed by an over-recovery has been reported previously \citep{lkg10}.


The hardness ratio increase shown in Figure 7 is interesting.
\citet{enm+10a} and \citet{bis+11} both reported on hard X-ray observations
made within 10 days after the 2009 outburst trigger.  They showed that in 
addition to a soft blackbody, a single power law of index $\Gamma \simeq 1.5$ could 
describe the data up to 200 keV.  The hardness ratio as measured by
{\it RXTE} increased significantly on MJD 54853, $\sim$5 days after the
{\it INTEGRAL} observation, at an epoch when
the 1--10~keV emission showed no remarkable change \citep{sk11}.  This suggests
that after MJD 54853, either a single power law no longer adequately described the data,
or the power-law index hardened.  This may indicate decoupling between the
soft and hard emission processes; hard X-ray monitoring after a future outburst
could check this.  Decoupling between the soft and hard X-ray emission mechanisms
would be interesting, as it would argue against a causal relationship in an
observed spectral correlation between these emission components \citep{kb10,enm+10b}.

In summary, we have reported on detailed {\it RXTE} and {\emph{Swift}} monitoring of the
AXP 1E 1547.0$-$5408 from 2008 October through 2009 December.  We have
presented pulsed flux time series in multiple energy bands, as well as
as a detailed analysis of the source's timing behavior, with phase-coherence being maintained only
for two $\sim$3~wk intervals following the 2008 and 2009 outbursts. We
showed that outside this time period several plausible timing
solutions fit the pulsar's frequency behavior in spite of relatively
dense monitoring.  We find a remarkable decrease in the
magnitude of $\dot{\nu}$ in concert with the falling of the pulsed
flux following the 2008 outburst, and note a similarity in the events
of the 2001-2002 outburst of 1E 1048.1$-$5937.  However this
interesting behavior is not fully replicated following the 2009
outburst of 1E 1547.0$-$5408.  We note a significant hardening of the
source $\sim$15 days after the 2009 outburst below 20~keV, at an epoch when the
1--10~keV X-ray flux shows no obvious change.  

The diversity of timing behaviors of
AXPs and magnetars in general near the epochs of radiative outbursts
continues to be challenging to understand, although the magnetar
framework first outlined by \citet{tlk02} and subsequently refined by
\citet{bel09} seems promising.

\acknowledgments 

We thank the {\it RXTE} and {\emph{Swift}} staff for their efficiency in scheduling the
above-described observations. VMK holds the Lorne Trottier Chair in Astrophysics
and Cosmology, and a
Canada Research Chair, a Killam Research Fellowship, and acknowledges
additional support from an NSERC Discovery Grant, from FQRNT via le Centre de
Recherche Astrophysique du Qu\'ebec and the Canadian Institute for
Advanced Research.


\clearpage
\begin{deluxetable}{cccccccccc}
\tabletypesize{\scriptsize}
\tablewidth{469.0pt}
\tablecaption
{{\em{RXTE}} Observations of 1E~1547.0$-$5408 Prior to 2009 January~22
\label{tableobs1}}
\tablehead
{                
\colhead{Obs.} &
\colhead{Obs.} &
\colhead{Date} &
\colhead{MJD} &
\colhead{Num.} &
\colhead{PCUs} &
\colhead{Obs.} &
\colhead{Usage\tablenotemark{c}} &
\colhead{Num.} &
\colhead{Profile} \\
\colhead{Num.} &
\colhead{ID} &
&
&
\colhead{of} &
\colhead{On\tablenotemark{a}} &
\colhead{Length\tablenotemark{b}} &
&
\colhead{of} &
\colhead{Code\tablenotemark{e}} \\
&
&
&
&
\colhead{Bursts} &
&
\colhead{(ks)} &
&
\colhead{TOAs\tablenotemark{d}} &
$\;$
}
\startdata
1 & D93017-10-01-00\tablenotemark{f} & 10/4/2008 & 54743.36 & 10$-$20 & 2,3 & 3.80 & t,p,f & 2 & A\\
2 & D93017-10-01-02 & 10/5/2008 & 54744.41 & 0 & 0,2 & 0.59 & t & 1 & $-$\\
3 & D93017-10-01-01 & 10/5/2008 & 54744.54 & 0 & 2 & 0.58 & t & 1 & $-$\\
4 & D93017-10-01-03 & 10/6/2008 & 54745.07 & 0$-$10 & 0,1,2,4 & 1.40 & t,p,f & 2 & B\\
5 & D93017-10-01-04 & 10/6/2008 & 54745.98 & 0 & 1,2 & 1.77 & t,p,f & 2 & C\\
6 & D93017-10-01-08 & 10/7/2008 & 54746.31 & 0$-$10 & 1,2 & 3.02 & t,p,fa & 2 & C\\
7 & D93017-10-01-09 & 10/7/2008 & 54746.37 & 0 & 2 & 3.51 & t,p,fa & 2 & C\\
8 & D93017-10-01-05 & 10/7/2008 & 54746.83 & 0 & 2 & 1.12 & t & 1 & $-$\\
9 & D93017-10-01-11\tablenotemark{g} & 10/8/2008 & 54747.03 & 0 & 0,2 & 2.19 & t,p,f & 2 & C\\
10 & D93017-10-01-06 & 10/8/2008 & 54747.64 & 0 & 2 & 1.01 & t & 1 & $-$\\
11 & D93017-10-01-10 & 10/9/2008 & 54748.92 & 0 & 2 & 1.65 & t,p,f & 1 & C\\
12 & D93017-10-02-00 & 10/10/2008 & 54749.60 & 0$-$10 & 2,3,4 & 1.12 & t,p,fa & 1 & C\\
13 & D93017-10-02-01 & 10/10/2008 & 54749.67 & 0 & 2,3p,4p & 1.63 & t,p,fa & 1 & C\\
14 & D93017-10-02-02 & 10/11/2008 & 54750.10 & 0 & 2 & 2.73 & t & 1 & $-$\\
15 & D93017-10-02-06 & 10/12/2008 & 54751.21 & 0 & 2,4 & 0.93 & t,p,f & 1 & C\\
16 & D93017-10-02-03\tablenotemark{h} & 10/13/2008 & 54752.13 & 0 & 0,2 & 2.38 & t,p,f & 2 & D\\
17 & D93017-10-02-04 & 10/14/2008 & 54753.37 & 0 & 1,2 & 3.46 & t,p,f & 2 & E\\
18 & D93017-10-02-05 & 10/16/2008 & 54755.33 & 0 & 1,2 & 3.49 & t,p,f & 2 & F\\
19 & D93017-10-03-00 & 10/17/2008 & 54756.31 & 0 & 2,4 & 3.46 & t,p,f & 2 & G\\
20 & D93017-10-03-01 & 10/18/2008 & 54757.58 & 0 & 0,2 & 1.10 & t,p,fa & 1 & H\\
21 & D93017-10-03-02 & 10/18/2008 & 54757.62 & 0 & 0,2 & 1.21 & t,p,fa & 1 & H\\
22 & D93017-10-04-00 & 10/20/2008 & 54759.00 & 0 & 0p,1p,2,4p & 7.61 & t,p,f & 2 & H\\
23 & D93017-10-04-01 & 10/21/2008 & 54760.96 & 0 & 2,4 & 3.49 & t,p,fa & 2 & H\\
24 & D93017-10-04-02 & 10/22/2008 & 54761.37 & 0 & 2 & 1.53 & t,p,fa & 1 & H\\
25 & D93017-10-04-03 & 10/23/2008 & 54762.20 & 0$-$10 & 2,3 & 3.52 & t,p,fa & 2 & H\\
26 & D93017-10-04-04 & 10/23/2008 & 54762.48 & 0 & 2,4 & 1.24 & t,p,fa & 1 & H\\
27 & D93017-10-05-00 & 10/24/2008 & 54763.18 & 0 & 2,4 & 3.50 & t,p,f & 2 & H\\
28 & D93017-10-05-02 & 10/25/2008 & 57764.84 & 0 & 2,3 & 1.05 & t,p,f & 1 & I\\
29 & D93017-10-05-03 & 10/27/2008 & 54766.06 & 0 & 2,4 & 2.06 & t,p,f & 2 & I\\
30 & D93017-10-05-01 & 10/29/2008 & 54768.80 & 0 & 2 & 3.04 & t,p,f & 2 & I\\
31 & D93017-10-06-00 & 11/1/2008 & 54771.81 & 0 & 2,4 & 3.92 & t,p,f & 2 & I\\
32 & D93017-10-06-01 & 11/5/2008 & 54775.02 & 0 & 1,2 & 3.69 & t,p,f & 2 & I\\
33 & D93017-10-06-03 & 11/6/2008 & 54776.98 & 0 & 2,4 & 1.66 & t,p,f & 2 & I\\
34 & D93017-10-07-00 & 11/9/2008 & 54779.99 & 0 & 2 & 1.65 & t,f & 2 & $-$\\
35 & D93017-10-07-01 & 11/13/2008 & 54783.07 & 0 & 2,4 & 1.48 & t,p,f & 2 & J\\
36 & D93017-10-08-00 & 11/16/2008 & 54786.24 & 0 & 2,3 & 2.08 & t,p,f & 2 & J\\
37 & D93017-10-08-01 & 11/20/2008 & 54790.08 & 0 & 2,4 & 3.21 & t,p,f & 2 & J\\
38 & D93017-10-09-00 & 11/23/2008 & 54793.09 & 0 & 2,3p & 6.26 & t,p,f & 2 & J\\
39 & D93017-10-10-00 & 12/3/2008 & 54803.00 & 0 & 1,2 & 1.95 & t,f & 2 & $-$\\
40 & D93017-10-10-01 & 12/3/2008 & 54803.10 & 0 & 2,3 & 2.92 & t,p,f & 2 & J\\
41 & D93017-10-10-02 & 12/3/2008 & 54803.20 & 0 & 1,2 & 1.92 & t & 2 & $-$\\
42 & D93017-10-11-00 & 12/9/2008 & 54809.90 & 0 & 2,4p & 6.97 & t,p,f & 2 & J\\
43 & D93017-10-12-00 & 12/19/2008 & 54819.60 & 0 & 0p,2 & 7.50 & t,p,f & 2 & K\\
44 & D93017-10-12-01 & 12/25/2008 & 54825.70 & 0 & 0p,1p,2 & 8.34 & t,p,f & 2 & K\\
45 & D93017-10-13-00 & 1/1/2009 & 54832.60 & 0 & 0,2 & 6.15 & t,p,f & 2 & K\\
46 & D93017-10-14-00 & 1/8/2009 & 54839.16 & 0 & 2 & 1.41 & t & 1 & $-$\\
47 & D93017-10-14-01 & 1/8/2009 & 58939.20 & 0 & 2 & 3.99 & t,p,f & 1 & K\\
48 & D93017-10-15-01 & 1/13/2009 & 54844.97 & 0 & 2,4 & 2.73 & t,p,f & 2 & K\\
49 & D93017-10-15-00 & 1/14/2009 & 54845.04 & 0 & 2,3 & 3.32 & t,p,f & 2 & K\\
\enddata
\tablenotetext{a}{If a PCU is on during one part of the observation only, 
its number is followed by a ``p.'' Data from all PCUs were used in the
timing analysis. Data from PCUs 2, 3, and~4 only were used in the pulsed
flux analysis.}
\tablenotetext{b}{Total time on source.}
\tablenotetext{c}{``t'' refers to this observation being used in the
timing analysis. ``p'' refers to this observation being used in the
pulse profile analysis. ``f'' refers to this observation being used
in the pulse flux analysis. ``fa'' indicates two consecutive
observations where the pulsed flux was averaged.}
\tablenotetext{d}{The number of independent TOAs extracted from this observation.}
\tablenotetext{e}{The letters in this column refer to the the time
intervals indicated in the bottom panel of Figure~\ref{f5}.}
\tablenotetext{f}{There was one {\emph{RXTE}} observation taken a few
hours prior to D93017-10-01-00, but it was in a data mode different
from all other obtained observations.}
\tablenotetext{g}{D93017-10-01-11 was originally named D93017-10-01-07.}
\tablenotetext{h}{The pulsed flux in this observation is significantly
higher than that in the previous observations.}
\end{deluxetable}

\clearpage
\begin{deluxetable}{cccccccccc}
\tabletypesize{\scriptsize}
\tablewidth{469.0pt}
\tablecaption
{{\em{RXTE}} Observations of 1E~1547.0$-$5408 from 2009 January~22 to
December~26
\label{tableobs2}}
\tablehead
{
\colhead{Obs.} &
\colhead{Obs.} &
\colhead{Date} &
\colhead{MJD} &
\colhead{Num.} &
\colhead{PCUs} &
\colhead{Obs.} &
\colhead{Usage\tablenotemark{c}} &
\colhead{Num.} &
\colhead{Profile} \\
\colhead{Num.} &
\colhead{ID} &
&
&
\colhead{of} &
\colhead{On\tablenotemark{a}} &
\colhead{Length\tablenotemark{b}} &
&
\colhead{of} &
\colhead{Code\tablenotemark{e}} \\
&
&
&
&
\colhead{Bursts} &
&
\colhead{(ks)} &
&
\colhead{TOAs\tablenotemark{d}} &
$\;$
}
\startdata
50 & D93017-10-17-00 & 1/22/2009 & 54853.86 & $>$20 & 2 & 6.66 & t,p,f & 2 & L\\
51 & D93017-10-16-01 & 1/23/2009 & 54854.78 & 10$-$20 & 2,3 & 2.64 & t,p,fa & 1 & L\\
52 & D93017-10-16-00 & 1/23/2009 & 54854.84 & $>$20 & 2,4 & 3.35 & t,p,fa & 2 & L\\
53 & D93017-10-16-05 & 1/24/2009 & 54855.82 & $>$20 & 2 & 10.29 & t,p,f & 2 & L\\
54 & D93017-10-16-02\tablenotemark{f} & 1/25/2009 & 54856.81 & $>$20 & 2 & 5.97 & t,p,f & 2 & L\\
55 & D94017-09-01-02 & 1/26/2009 & 54857.15 & $>$20 & 2,4 & 1.81 & t,p & 2 & L\\
56 & D94017-09-01-01 & 1/28/2009 & 54859.88 & $>$20 & 1p,2,3p & 10.32 & t,p,f & 2 & L\\
57 & D94017-09-01-03 & 1/29/2009 & 54860.74 & $>$20 & 0p,1,2,4p & 5.58 & t,p,f & 2 & L\\
58 & D94017-09-02-01 & 1/31/2009 & 54862.56 & 10$-$20 & 0,2 & 1.76 & t,p,fa & 1 & L\\
59 & D94017-09-02-00 & 1/31/2009 & 54862.69 & 10$-$20 & 0p,1p,2,4 & 17.46 & t,p,fa & 2 & L\\
60 & D94017-09-02-02 & 2/1/2009 & 54863.75 & 10$-$20 & 1p,2 & 5.91 & t,p,f & 2 & L\\
61 & D94017-09-02-03 & 2/1/2009 & 54863.87 & $>$20 & 2,4 & 3.27 & t,p,fa & 2 & L\\
62 & D94017-09-02-04 & 2/1/2009 & 54863.93 & $>$20 & 0,2 & 3.26 & t,p,fa & 2 & L\\
63 & D94017-09-03-00 & 2/6/2009 & 54868.72 & 10$-$20 & 1p,2,3p & 5.81 & t,p,fa & 2 & M\\
64 & D94017-09-03-01 & 2/6/2009 & 54868.83 & 0$-$10 & 2,4 & 2.31 & t,p,fa & 1 & M\\
65 & D94017-09-04-00 & 2/13/2009 & 54875.58 & 10$-$20 & 2,3p,4p & 7.61 & t,p,f & 2 & M\\
66 & D94017-09-04-01 & 2/18/2009 & 54880.7 & 0$-$10 & 1p,2 & 5.12 & t,p,f & 2 & M\\
67 & D94017-09-05-00 & 2/28/2009 & 54890.1 & 0$-$10 & 1,2 & 4.12 & t,p,fa & 2 & M\\
68 & D94017-09-05-01 & 2/28/2009 & 54890.17 & 0$-$10 & 2,3 & 1.58 & t,p,fa & 1 & M\\
69 & D94017-09-06-00 & 3/6/2009 & 54896.84 & 0 & 0p,1p,2 & 7.24 & t,p,f & 2 & M\\
70 & D94017-09-07-02 & 3/16/2009 & 54906.45 & 0 & 1,2 & 1.66 & t,p,fa & 2 & M\\
71 & D94017-09-07-01 & 3/16/2009 & 54906.52 & 0$-$10 & 2,4 & 2.01 & t,p,fa & 2 & M\\
72 & D94017-09-07-00 & 3/16/2009 & 54906.59 & 0 & 1,2 & 2.37 & t,p,f & 2 & M\\
73 & D94017-09-08-00 & 3/21/2009 & 54911.76 & 10$-$20 & 0p,1p,2 & 5.87 & t,p,f & 2 & M\\
74 & D94017-09-09-01 & 3/30/2009 & 54920.53 & 0 & 2 & 2.46 & t,p,f & 2 & M\\
75 & D94017-09-09-00\tablenotemark{g} & 3/30/2009 & 54920.59 & $>$20 & 1,2 & 3.06 & t,p,f & 2 & M\\
76 & D94017-09-10-00 & 4/5/2009 & 54926.54 & 0$-$10 & 0p,1p,2 & 5.84 & t,p,f & 2 & M\\
77 & D94017-09-11-00 & 4/14/2009 & 54935.25 & 0 & 2,4 & 5.98 & t,p,f & 2 & M\\
78 & D94017-09-12-00 & 4/19/2009 & 54940.87 & 0$-$10 & 1p,2,4p & 6.54 & t,p,f & 2 & N\\
79 & D94017-09-13-00 & 4/25/2009 & 54946.63 & 0 & 2,3 & 2.98 & t,p,fa & 2 & N\\
80 & D94017-09-13-01 & 4/25/2009 & 54946.7 & 10$-$20 & 2,4 & 2.97 & t,p,fa & 2 & N\\
81 & D94017-09-14-00 & 5/3/2009 & 54954.8 & 0 & 2,4p & 6.98 & t,p,f & 2 & N\\
82 & D94017-09-15-00 & 5/11/2009 & 54962.52 & 0 & 2,4p & 5.95 & t,p,f & 2 & N\\
83 & D94017-09-16-00 & 5/18/2009 & 54969.26 & 0 & 2,3 & 2.91 & t,p,fa & 2 & N\\
84 & D94017-09-16-01 & 5/18/2009 & 54969.32 & 0 & 2,4 & 3.03 & t,p,fa & 2 & N\\
85 & D94017-09-17-00 & 5/26/2009 & 54977.44 & 0 & 2,3p,4p & 6.12 & t,p,f & 2 & N\\
86 & D94017-09-18-00 & 6/4/2009 & 54986.19 & 0 & 2,4p & 5.95 & t,p,f & 2 & N\\
87 & D94017-09-19-02 & 6/9/2009 & 54991.83 & 0 & 2,3 & 1.48 & t,p,f & 2 & N\\
88 & D94017-09-19-00 & 6/10/2009 & 54992.02 & 0 & 2,3 & 2.97 & t,p,fa & 2 & N\\
89 & D94017-09-19-01 & 6/10/2009 & 54992.09 & 0 & 2 & 1.47 & t,p,fa & 1 & N\\
90 & D94017-09-20-00 & 6/17/2009 & 54999.54 & 0 & 2,3p,4p & 7.14 & t,p,f & 2 & N\\
91 & D94017-09-21-00 & 6/22/2009 & 55004.6 & 0 & 2,4 & 5.6 & t,p,f & 2 & N\\
92 & D94427-01-01-00 & 6/30/2009 & 55012.43 & 0 & 2 & 5.93 & t,p,f & 2 & N\\
93 & D94427-01-02-00 & 7/6/2009 & 55018.26 & 0 & 2,3p,4p & 5.89 & t,p,f & 2 & O\\
94 & D94427-01-03-00 & 7/14/2009 & 55026.02 & 0 & 2,3 & 4.12 & t,p,f & 2 & O\\
95 & D94427-01-04-00 & 7/23/2009 & 55035.66 & 0$-$10 & 2,3p,4p & 6.18 & t,p,f & 2 & O\\
96 & D94427-01-05-00 & 7/31/2009 & 55043.44 & 0 & 0p,1p,2 & 7.32 & t,p,f & 2 & O\\
97 & D94427-01-06-00 & 8/11/2009 & 55054.23 & 0 & 0p,2,3p & 5.52 & t,p,f & 2 & O\\
98 & D94427-01-07-00 & 8/20/2009 & 55063.2 & 0 & 1,2 & 1.89 & t,p,fa & 2 & O\\
99 & D94427-01-07-01 & 8/20/2009 & 55063.25 & 0$-$10 & 2,3 & 3.21 & t,p,fa & 2 & O\\
100 & D94427-01-08-00 & 8/28/20009 & 55071.1 & 0 & 0p,2 & 3.35 & t,p,fa & 2 & O\\
101 & D94427-01-08-01 & 8/28/2009 & 55071.16 & 0 & 2 & 3.63 & t,p,fa & 2 & O\\
102 & D94427-01-09-00 & 9/6/2009 & 55080.99 & 0 & 2 & 7.72 & t,p,f & 2 & O\\
103 & D94427-01-13-00 & 10/8/2009 & 55112.11 & 0 & 0p,1p,2 & 5.94 & t,p,f & 2 & P\\
104 & D94427-01-14-00 & 10/16/2009 & 55120.94 & 0 & 2,3 & 3.63 & t,p,f & 2 & P\\
105 & D94427-01-14-01 & 10/17/2009 & 55121.02 & 0 & 2,4 & 2.31 & t,p,f & 2 & P\\
106 & D94427-01-15-00 & 11/3/2009 & 55138.14 & 0 & 0p,1p,2 & 5.95 & t,p,f & 2 & P\\
107 & D94427-01-16-00 & 11/11/2009 & 55146.14 & 0 & 0p,1p,2 & 5.95 & t,p,f & 2 & P\\
108 & D94427-01-10-00 & 11/20/2009 & 55155.22 & 0 & 2,3 & 5.93 & t,p,f & 2 & P\\
109 & D94427-01-17-00 & 11/27/2009 & 55162.88 & 0 & 2 & 5.95 & t,p,f & 2 & P\\
110 & D94427-01-11-00 & 12/5/2009 & 55170.92 & 0 & 2 & 6.24 & t,p,f & 2 & P\\
111 & D94427-01-18-01 & 12/12/2009 & 55177.92 & 0 & 2,4 & 2.43 & t,p,fa & 2 & P\\
112 & D94427-01-18-00 & 12/13/2009 & 55178.04 & 0 & 2 & 2.98 & t,p,fa & 1 & P\\
113 & D94427-01-12-00 & 12/19/2009 & 55184.93 & 0 & 1p,2,4p & 5.18 & t,p,f & 2 & P\\
114 & D94427-01-19-00 & 12/26/2009 & 55191.12 & 0 & 2,3p,4p & 6.38 & t,p,f & 2 & P\\
\enddata
\tablenotetext{a}{If a PCU is on during one part of the observation only,
its number is followed by a ``p.'' Data from all PCUs were used in the
timing analysis. Data from PCUs 2, 3, and~4 only were used in the pulsed
flux analysis.}
\tablenotetext{b}{Total time on source.}
\tablenotetext{c}{``t'' refers to this observation being used in the
timing analysis. ``p'' refers to this observation being used in the
pulse profile analysis. ``f'' refers to this observation being used
in the pulse flux analysis. ``fa'' indicates two consecutive
observations where the pulsed flux was averaged.}
\tablenotetext{d}{The number of independent TOAs extracted from this observation.}
\tablenotetext{e}{The letters in this column refer to the time
intervals indicated in the bottom panel of Figure~\ref{f5}.}
\tablenotetext{f}{ID D93017-10-16-02 is also called D94017-09-01-00.}
\tablenotetext{g}{In observation D94017-09-09-00, the count rate is
unusually high, and one burst has a particularly long tail.}
\end{deluxetable}

\clearpage
\begin{deluxetable}{lcccc}
\tabletypesize{\scriptsize}
\tablewidth{290.0pt}
\tablecaption
{
{\emph{Swift}} Observations of 1E~1547.0$-$5408 from 2008 October to
2009 December Used in this Paper
\label{table3}
}
\tablehead
{
Obs. ID &
Date &
MJD &
Obs. Length\tablenotemark{a} &
Phase- \\
&
&
&
(ks)&
Connected\tablenotemark{b}
}
\startdata
00330353001  & 2008-10-03 & 54742.80 & 14.9 & yes \\
00330353002  & 2008-10-04 & 54743.79 & 4.9 & yes \\
00330353004  & 2008-10-05 & 54745.14 & 10.5 & yes \\
00330353005  & 2008-10-07 & 54746.51 & 7.7 & yes \\
00330353006  & 2008-10-08 & 54747.11 & 4.5 & yes \\
00330353007  & 2008-10-09 & 54748.39 & 3.6 & yes \\
00330353008  & 2008-10-10 & 54749.46 & 3.8 & yes \\
00330353010  & 2008-10-12 & 54751.53 & 3.9 & yes \\
00330353011  & 2008-10-13 & 54752.40 & 3.4 & yes \\
00330353012  & 2008-10-16 & 54755.14 & 4.0 & yes \\
00330353013  & 2008-10-17 & 54757.38 & 5.0 & yes \\
00330353014  & 2008-10-20 & 54759.80 & 3.9 & yes \\
00330353015  & 2008-10-22 & 54761.43 & 3.8 & yes \\
00330353016  & 2008-10-24 & 54763.21 & 3.5 & yes \\
00090007024  & 2009-01-04 & 54835.13 & 3.3 & no \\
00090007025  & 2009-01-12 & 54844.13 & 4.2 & no \\
00340573000  & 2009-01-22 & 54853.20 & 7.6 & yes \\
00340573001  & 2009-01-22 & 54853.55 & 9.4 & yes \\
00090007026  & 2009-01-23 & 54855.05 & 8.2 & yes \\
00090007027  & 2009-01-25 & 54856.19 & 3.2 & yes \\
00090007028  & 2009-01-26 & 54857.25 & 3.5 & yes \\
00090007035  & 2009-02-02 & 54864.72 & 4.1 & yes \\
00090956040  & 2009-02-05 & 54867.57 & 6.1 & yes \\
00090007036  & 2009-02-12 & 54874.32 & 6.0 & yes \\
00090007037  & 2009-02-22 & 54884.63 & 4.6 & no \\
00090007038  & 2009-03-04 & 54894.61 & 3.9 & no \\
00090007039  & 2009-03-13 & 54904.45 & 4.0 & no \\
00090007040  & 2009-03-24 & 54914.90 & 4.1 & no \\
00030956043  & 2009-04-29 & 54950.47 & 1.6 & no \\
00030956045  & 2009-05-27 & 54978.52 & 2.1 & no \\
00030956053  & 2009-09-16 & 55090.52 & 1.4 & no \\
00030956054  & 2009-09-30 & 55104.63 & 3.3 & no \\
00030956055  & 2009-10-14 & 55118.09 & 1.9 & no \\
\enddata
\tablenotetext{a} {XRT time on source.}
\tablenotetext{b} {``yes'' means that this observation
belongs to a period of time covered by a unique phase-connected
timing solution reported in this paper. See Section~\ref{sec:timing} in the text.}
\end{deluxetable}

\clearpage
\begin{deluxetable}{lcc}
\tabletypesize{\small}
\tablewidth{315.0pt}
\tablecaption
{
Spin Parameters for 1E~1547.0$-$5408\tablenotemark{a}$\;$ for the
Phase-Connected Segments
\label{table4}
}
\tablehead
{
&
\colhead{Post-2008 outburst} &
\colhead{Post-2009 outburst} \\
\colhead{Parameter} &
\colhead{ephemeris spanning} &
\colhead{ephemeris spanning} \\
&
\colhead{29 days} &
\colhead{23 days}
}
\startdata
MJD range & 54743.4$-$54771.8 & 54853.9$-$54875.7 \\
No. TOAs\tablenotemark{b} & 46~{\emph{RXTE}} + 14~{\emph{Swift}} & 31~{\emph{RXTE}} + 8~{\emph{Swift}} \\
$\nu$ (Hz) & 0.48277893(4) & 0.48259625(3)  \\
$\dot{\nu}$ (10$^{-12}$ Hz s$^{-1}$) & $-$6.19(8) & $-$5.21(4) \\
$\ddot{\nu}$ (10$^{-18}$ Hz s$^{-2}$) & $-$6.69(7) & $<$~1.1\tablenotemark{c} \\
Epoch (MJD) & 54743.000 & 54854.000 \\
RMS residual & 3\% & 4\% \\
\enddata
\tablenotetext{a} {Numbers in parentheses are TEMPO-reported 1$\sigma$
uncertainties.} 
\tablenotetext{b} {The number of {\emph{RXTE}} TOAs is larger than the number of
{\emph{RXTE}} observations because the longest observations yielded multiple TOAs.}
\tablenotetext{c} {3$\sigma$ upper limit.}
\end{deluxetable}


\clearpage

\clearpage

\clearpage

\clearpage

\clearpage


\clearpage

\clearpage
\clearpage

\clearpage
\begin{figure}
\includegraphics[scale=.8]{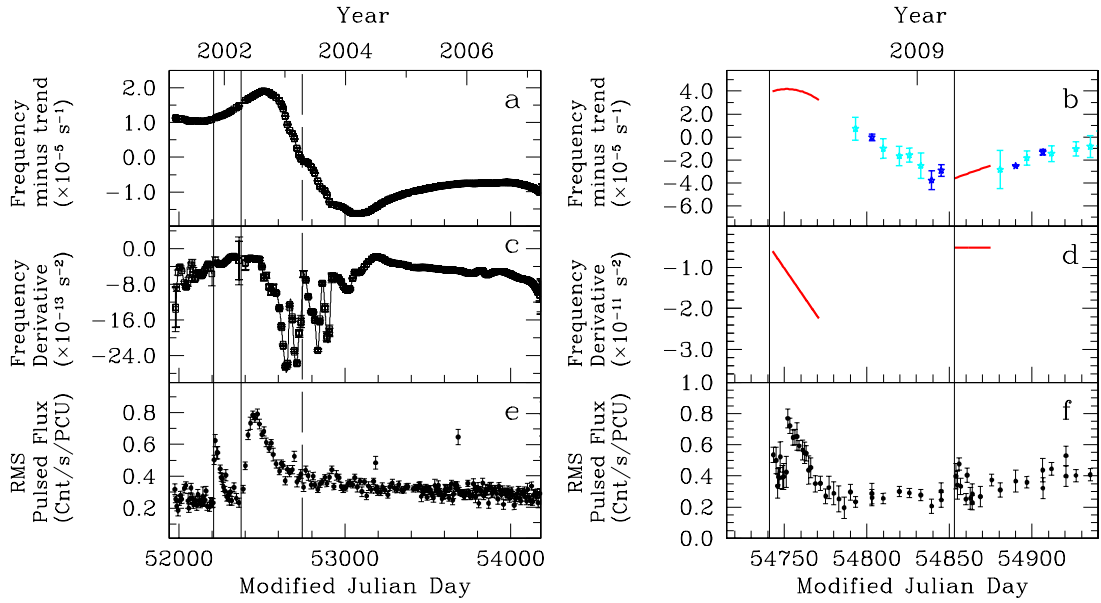}
\caption{
Comparison of the timing and pulsed flux properties of
1E~1048.1$-$5937 (left column) and 1E~1547.0$-$5408 (right column).
(a) Frequency as a function of time for 1E~1048.1$-$5937 with a linear
trend subtracted (for details, see Figures~3 and~15 of
\citeauthor{dkg09}~\citeyear{dkg09}).
(b) Frequency as a function of time for 1E~1547.0$-$5408 with a linear
trend subtracted (see Figure~2 for details).
(c) Frequency derivative as a function of time for 1E~1048.1$-$5937.
(d) Frequency derivative as a function of time for 1E~1547.0$-$5408.
(e) Pulsed flux as a function of time in the 2$-$10~keV band for
1E~1048.1$-$5937.
(f) Pulsed flux as a function of time in the 2$-$10~keV band for
1E~1547.0$-$5408.
\label{f8}}
\end{figure}
\clearpage


\bibliographystyle{apj}

\end{document}